\documentclass[showpacs,preprintnumbers,twocolumn,
amsmath,amssymb,groupedaddress,superscriptaddress]{revtex4}
\usepackage{graphicx}
\usepackage{dcolumn}
\usepackage{bm}

\begin{document}

\preprint{}

\title{Longitudinal and Transverse Scaling Functions within the Coherent Density Fluctuation Model}

\author{A.N. Antonov}
\affiliation{Institute for Nuclear Research and Nuclear Energy,
Bulgarian Academy of Sciences, Sofia 1784, Bulgaria}

\author{M.V. Ivanov}
\affiliation{Institute for Nuclear Research and Nuclear Energy,
Bulgarian Academy of Sciences, Sofia 1784, Bulgaria}

\author{M.B.~Barbaro}
\affiliation{Dipartimento di Fisica Teorica, Universit\`a di Torino
and INFN, Sezione di Torino, Torino, Italy}

\author{J.A.~Caballero}
\affiliation{Departamento de F\'\i sica At\'omica, Molecular y
Nuclear, Universidad de Sevilla, E-41080 Sevilla, Spain}

\author{E. Moya de Guerra}
\affiliation{Instituto de Estructura de la Materia, CSIC, Serrano
123, E-28006 Madrid, Spain}

\affiliation{Departamento de Fisica Atomica, Molecular y Nuclear,
Facultad de Ciencias Fisicas, Universidad Complutense de Madrid,
E-28040 Madrid, Spain}

%\date{\today}

\begin{abstract}

We extend our previous description of the superscaling phenomenon in inclusive electron scattering within the Coherent Density Fluctuation Model (CDFM). This model is a natural extension to finite nuclei of the Relativistic Fermi Gas Model (RFG) within which the scaling variable $\psi^{\prime}$ was introduced. In this work we propose a new modified CDFM approach to calculate the total, longitudinal and transverse scaling functions built up from the hadronic tensor and the longitudinal and transverse response functions in the RFG. We test the superscaling behavior of the new CDFM scaling functions by calculating the cross sections of electron scattering (in QE- and $\Delta$-region for nuclei with $12\leq A\leq208$ at different energies and angles) and comparing to available experimental data. The new modified CDFM approach is extended to calculate charge-changing neutrino and antineutrino scattering on $^{12}$C at 1~GeV incident energy. %at energies from 1 to 2~GeV not only in the quasielastic but also in the delta excitation region. The scaling function obtained within the CDFM is applied to neutral current neutrino and antineutrino scattering with energies of 1~GeV from $^{12}$C with a proton and neutron knockout.

\end{abstract}


\maketitle

\section[]{INTRODUCTION\label{sect1ant}}

The Relativistic Fermi Gas model in~\cite{ant01,ant02} has been the basis to define the scaling variable $\psi^\prime$ and to introduce the first theoretical considerations of the superscaling phenomenon. Superscaling has been observed in inclusive electron scattering from nuclei (see, \emph{e.g.}~\cite{ant01,ant02,ant03,ant04}). The term ``superscaling'' includes scaling of the first and the second kind (independence of the reduced cross section on the momentum transfer $q$ and the mass number $A$, respectively) that have been seen at excitation energies below the quasielastic (QE) peak. These studies can be considered a part of more general investigations that follow the ideas of West~\cite{ant05} on scaling phenomena. They include also the studies of the related $y$-scaling in high-energy electron-nuclei scattering (\emph{e.g.} \cite{ant05,ant06,ant07,ant08,ant09,ant10,ant11,ant12,ant13}). It has been shown in both $y$- and $\psi^\prime$- scaling analyses that the scaling function is sensitive to the high-momentum components of the spectral function and, consequently, to the tail of the nucleon momentum distributions $n(k)$. Thus, the knowledge of the scaling function can provide important information about the dynamical ground-state properties of the nuclei. In the $\psi^\prime< -1$ region, superscaling is due to the specific high-momentum tail of $n(k)$ caused by short-range and tensor correlations, which is similar for all nuclei and which is in turn related to specific properties of the nucleon-nucleon (NN) forces near the repulsive core. Even more, it has been shown in~\cite{ant14} that the behavior of the scaling function $f(\psi^\prime)$ for values of $\psi^\prime < -1$ depends on the particular form of the power-law asymptotic of $n(k)$ at large $k$ related to a corresponding  behavior of the in-medium NN forces around the core. The latter dependence together with the existing link between the asymptotic behavior of $n(k)$ and the NN forces make it possible to conclude that the inclusive QE electron scattering at $\psi'\lesssim -1$ provides important information on the NN forces in the nuclear medium.

Confirming the superscaling behavior of the world data on inclusive electron scattering, the analyses in~\cite{ant03,ant04} have shown the necessity to consider this phenomenon on the basis of more complex dynamical picture of realistic finite nuclear systems beyond the RFG. Indeed, the scaling function in the RFG model is $f(\psi^\prime) = 0$ for $\psi^\prime\leq  - 1$, whereas the experimental scaling function extracted from $(e,e')$ data extends up to $\psi^\prime\approx -2$, where the effects beyond the mean-field approximation become important. One of such approaches is the Coherent Density Fluctuation Model (\emph{e.g.}~\cite{ant15,ant16}) that is a natural extension of the Fermi gas model based on the generator coordinate method~\cite{ant17} and includes long-range correlations (LRC) of collective type. The QE scaling function $f(\psi^\prime)$ is deduced in the CDFM on the basis of realistic density and momentum distributions in nuclei and it agrees with the empirical data for the scaling function for negative values of $\psi^\prime \leq- 1$~\cite{ant18,ant19,ant20,ant21,ant22}. This agreement is related to the realistic high-momentum tail of the nucleon momentum distribution in the CDFM, that is similar for a wide range of nuclei, in contrast with the sharp behavior of $n(k)$ as a function of $k$ in the RFG (see, \emph{e.g.} Fig.~3 of~\cite{ant18}, Fig.~2 of~\cite{ant19} and the analysis in~\cite{ant14}). A number of studies of the superscaling has been published in the last years (\emph{e.g.}~\cite{ant23,ant24,ant25,ant26,ant27,ant28,ant29,ant30,ant31,ant32,ant33,ant34,ant35}). A ``universal'' phenomenological  QE scaling function $f^\text{QE}(\psi^\prime)$ has been obtained~\cite{ant03,ant04,ant23,ant24} on the basis of the available separation of inclusive electron scattering data into their longitudinal and transverse contributions for nuclei with $A>4$. An unexpected feature of this scaling function extracted from the superscaling analyses (SuSA) is its asymmetric shape with respect to the peak position $\psi^\prime=0$ with a long tail extended towards positive $\psi^\prime$ values. This is in contrast to the RFG scaling function that is symmetric. Detailed studies of this asymmetry~\cite{ant36,ant37,ant38} within the relativistic mean-field (RMF) approach have shown the important role played by a proper description of final-state interactions (FSI) to reproduce the asymmetric shape of  $f(\psi^\prime)$. The existence of the asymmetric tail of the scaling function has also been shown recently in~\cite{ant35} in a model accounting for pairing BCS-type correlations.
% in the nuclear ground state and for final states consisting of a single plane-wave nucleon plus a BCS recoiling daughter nucleus.

The approach of SuSA to the quasielastic electron scattering has been extended in~\cite{ant24}, to the $\Delta$-resonance excitation region.

The features of superscaling in inclusive electron-nucleus scattering have made it possible to initiate analyses of neutrino and antineutrino scattering off nuclei on the same basis (\emph{e.g.}~\cite{ant24,ant36,ant39,ant40}). Neutrino- (antineutri\-no-) nucleus charge-changing (CC)~\cite{ant36} or neutral-current (NC)~\cite{ant32} scattering cross sections for intermediate to high energies can be calculated by multiplying the elementary single-nucleon CC or NC neutrino (antineutrino) cross sections by the corresponding scaling functions. The assumptions leading to such a procedure have been tested within the RMF plus FSI model~\cite{ant41}. Here we mention a number of other theoretical studies of CC (see \emph{e.g.} Refs.~\cite{ant42,ant43,ant44,ant45,ant46,ant47,ant48,ant49,ant50,ant51}) and NC (\emph{e.g.}~\cite{ant42,ant43,ant50,ant52,ant53,ant54,ant55,ant56}) neutrino- and antineutrino- nucleus scattering in the last years.

The CDFM scaling function has been used to predict cross sections for several processes: inclusive electron scattering in the QE and $\Delta$- regions~\cite{ant21,ant22} and neutrino (antineutrino) scattering both for CC~\cite{ant22} and for NC~\cite{ant56} processes (u-channel inclusive processes).
%The scaling function in the CDFM has been applied to calculate cross sections of inclusive electron scattering, as well as of CC neutrino (antineutrino) scattering in the QE region on $^{12}$C with energies of the incident particles from 1 to 2~GeV in~\cite{ant21}. The asymmetry of the scaling function has been introduced in a phenomenological way. This function has been applied to calculate and predict CC neutrino (antineutrino) scattering on the same nucleus and at the same energies also in the $\Delta$-excitation region in~\cite{ant22}. The CDFM scaling function has been also applied to calculate NC neutrino (antineutrino) scattering with 1~GeV energy from $^{12}$C with a proton and neutron knockout (u-channel inclusive processes) in~\cite{ant56}. It has been shown by these analyses that the constructed CDFM scaling function (that is based on realistic density and momentum distributions) is an essential ingredient of the approach for the calculations of cross sections of electron and neutrino (antineutrino) scattering off nuclei.
%It has been pointed out that the CDFM is useful to explore both the u- and t- channel scaling criteria.
The CDFM analyses became useful to obtain information about the role of the nucleon momentum and density distributions for the explanation of superscaling in lepton-nucleus scattering~\cite{ant19,ant21}. It may also prove to be useful to explore the extension of the u- and t- channel scaling criteria beyond independent particle models.

It is important to point out that the physics contained in the ``experimental'' scaling function comes not only from the initial but also from the final states involved in the scattering process. Hence, caution should be placed on the general connection between the scaling function and the spectral function (or momentum distribution). Nevertheless, following the general consideration introduced in~\cite{ant14} on the relationship between $f(\psi')$ and the nucleon momentum distribution, $n(k)$, it was found within the CDFM~\cite{ant19} that the slope of the QE scaling function $f(\psi^\prime)$ at negative values of $\psi^\prime$ crucially depends on the high-momentum tail of the momentum distribution $n(k)$ at larger values of $k$ ($k > 1.5$~fm$^{-1}$). Moreover, the sensitivity of the scaling function in the CDFM to the particular behavior of $n(k)$ in different regions of $k$ has been studied in~\cite{ant19}, showing that the available empirical data on $f(\psi^\prime)$ are informative for $n(k)$ for momentum $k$ up to $k \leq 2$--$2.5$~fm$^{-1}$.

In our previous works~\cite{ant18,ant19,ant20,ant21} we obtained the CDFM scaling function $f(\psi^\prime)$ starting from the RFG model scaling function $f_\text{RFG}(\psi^\prime)$ and convoluting it with the weight function $|F(x)|^2$ that is related equivalently to either the density $\rho(r)$ or the momentum distribution $n(k)$ in nuclei. Thus, the CDFM scaling function is an infinite superposition of weighted RFG scaling functions. This approach improves upon RFG and enables one to describe the scaling function for realistic finite nuclear systems. In the approach in~\cite{ant18,ant19,ant20,ant21} the longitudinal and transverse scaling functions are equal $f_L(\psi^\prime) =  f_T(\psi^\prime)$ (this is the so-called scaling of zero-kind) that is also a property of the RFG scaling functions. The aim of this work is to develop a new CDFM approach in which we start directly from the hadronic RFG~\cite{ant01} tensor $W^{\mu\nu}$ and the corresponding response functions $R_{L,T}$, and convolute them with the CDFM function $|F(x)|^2$. We call this new approach CDFM$_\text{II}$ to distinguish it from our former version to which we refer as CDFM$_\text{I}$. This method provides a more general way to apply CDFM ideas and to go beyond RFG in the construction of scaling functions. Particularly it allows us to study the possible violation of the zero-kind scaling ($f_L(\psi^\prime)$ not equal to $f_T(\psi^\prime)$) and to compare the behavior of $f_L$ and $f_T$ to that from other approaches (\emph{e.g.} Relativistic Plane-Wave Impulse Approximation (RPWIA)). It can be seen in our work that the CDFM$_\text{II}$ scaling function calculated for different values of the transferred momentum $q$ shows both a saturation of its asymptotic behavior and also the region of appearance of the scaling of the first kind (at values of the transverse momentum of the order or higher than $0.5$~GeV/c). The main difference between CDFM$_\text{I}$ and CDFM$_\text{II}$ is that in CDFM$_\text{I}$ the LRC are taken ``a posteriori'' in the scaling function, once the RFG scaling function has been derived from the total inclusive cross section, while in the CDFM$_\text{II}$ the correlations are included through the weighting function $|F(x)|^2$ in the hadronic tensor, \emph{i.e.} they are included at an earlier stage in the derivation of the cross section. This allows us to study the emergence of scaling within the model, as well as possible differences between longitudinal and transverse scaling functions.
%One more important feature is that it is possible to calculate the CDFM scaling functions $f(\psi^\prime)$, $f_L(\psi^\prime)$ and $f_T(\psi^\prime)$ at any values of $q$, not only at $q > 2p_F$ but also at $q< 2p_F$ ($p_F$ being the Fermi momentum) where the scaling phenomenon does not hold in the RFG model on which the CDFM is based.

The second aim of the present work is to apply the obtained CDFM$_\text{II}$ scaling functions ($f$, $f_L$ and $f_T$)  to calculate  cross sections of inclusive electron scattering off various nuclei, as well as cross sections of CC neutrino (antineutrino) scattering on $^{12}$C at intermediate energies.

The theoretical scheme used in the present work is given in Sec.~\ref{sect2ant}. It includes the basic relationships of the RFG  model for the hadronic tensor, the response functions, as well as the procedure to obtain the CDFM$_\text{II}$ scaling functions. The results for $f(\psi^\prime)$, $f_L(\psi^\prime)$ and $f_T(\psi^\prime)$, as well as those from calculations of inclusive electron scattering cross sections in both (CDFM$_\text{I}$ and CDFM$_\text{II}$) approaches and of cross sections of CC neutrino reactions on $^{12}$C are presented and discussed in Sec.~\ref{sect3ant}. The conclusions are summarized in Sec.~\ref{sect4ant}.

\section[]{THEORETICAL SCHEME\label{sect2ant}}

We begin this section with a brief discussion of the basic
formalism for inclusive electron scattering from
nuclei~\cite{ant01} in which an electron with four-momentum
${K}^\mu=(\epsilon,\mathbf{k})$ is scattered through an angle
$\theta=\measuredangle(\mathbf{k},\mathbf{k}')$ to four-momentum
${K}'^\mu=(\epsilon',\mathbf{k}')$. The four-momentum
transferred in the process is then
${Q}^\mu=({K}-{K}')^\mu=(\omega,\mathbf{q})$,
where $\omega =\epsilon -\epsilon ^{\prime }$, ${q}=|\mathbf{q}|
=\mathbf{k}-\mathbf{k}^{\prime }$, and $Q^2=\omega^2-q^2\leq0$. In the relativistic limit (ERL) $|\mathbf{k}|\cong\epsilon\gg m_e$ and
$|\mathbf{k}'|\cong\epsilon'\gg m_e$, where $m_e$ is the electron
mass. In the one-photon-exchange approximation, the
double-differential cross section in the laboratory system can be
written in the form
\begin{multline}
    \dfrac{d^2\sigma}{d\Omega
d\epsilon'}=\sigma_M\bigg[\left[\dfrac{Q^2}{q^2}\right]^2R_L(q,\omega)+\\+\left[
\dfrac{1}{2}\left|\dfrac{Q^2}{q^2}\right|+\tan^2\dfrac{\theta}{2}\right]
R_T(q,\omega)\bigg]\label{s2e1},
\end{multline}
where $L$ ($T$) refer to responses with longitudinal (transverse)
projections (\emph{i.e.}, with respect to the momentum transfer
direction) of the nuclear currents, and where the Mott cross section
is given by
\begin{eqnarray}
\sigma_M=\left[\dfrac{\alpha\cos(\theta /2)}{2\epsilon\sin^2(\theta
/2)}\right]^2,\label{s2e2}
\end{eqnarray}
with $\alpha$ the fine-structure constant.

This cross section is obtained by contracting leptonic and hadronic
current-current interaction electromagnetic tensors and hence it is
proportional to $\eta_{\mu\nu}W^{\mu\nu}$. The leptonic tensor $\eta_{\mu\nu}$ may
be calculated in the standard way involving traces of Dirac $\gamma$
matrices and under ERL conditions becomes
\begin{eqnarray}
\eta_{\mu\nu}={K}_\mu{K}'_\nu+{K}'_\mu{K}_\nu-g_{\mu\nu}{K}\cdot{K}'.\label{s2e3}
\end{eqnarray}
Contracting this with a general hadronic tensor $W^{\mu\nu}$ and
rewriting the cross section in Eq.~(\ref{s2e1}), we have the
following for the two response functions (summation convention on
repeated indices):
\begin{gather}
R_L(q,\omega)=W^{00},\label{s2e4}\\
R_T(q,\omega)=
-\left(g_{ij}+\dfrac{q_iq_j}{q^2}\right)W^{ij}.\label{s2e5}
\end{gather}

In the RFG model the hadronic tensor $W^{\mu\nu}$ can be expressed by:
\begin{align}
W^{\mu\nu}\!=&\!\dfrac{3\mathcal{N}m_N^2}{4\pi p^3_F}\!\!\int\limits{}{}\!\!\dfrac{d^3p}{E(\mathbf{p})E(\mathbf{p}+\mathbf{q})}
\theta(p_F-|\mathbf{p}|\!)\notag\\
&\times\theta(|\mathbf{p}+\mathbf{q}|-p_F\!)\delta[\omega-[E(\mathbf{p}+\mathbf{q})-E(\mathbf{p})]]\notag\\
&\times f^{\mu\nu}(P+Q,P),~\mathcal{N}=N,Z\label{s2e6}
\end{align}
where the scattering is assumed to involve a struck nucleon of mass $m_N$, four-momentum $P = [E(\mathbf{p}),\mathbf{p}]$ with corresponding (on-shell) energy $E(\mathbf{p}) = {(\mathbf{p}^2+m_N^2)}^{1/2}$ lying below the Fermi momentum $p_F$ and, having supplied energy and momentum $\omega$ and $q$, respectively, to the nucleon, resulting in a four-momentum $(P + Q)^\mu$ lying above the Fermi surface. $f^{\mu\nu}(P+Q,P)$ is the single-nucleon response tensor
%obtained by Lorentz transforming the measured response involving the
%system where the struck nucleon is at rest to
evaluated in the system where the struck nucleon has four-momentum $P$:
\begin{eqnarray}\label{s2e7}
f^{\mu\nu}(P+Q,P)=-W_1(\tau)\left(g^{\mu\nu}-\dfrac{Q^\mu
Q^\nu}{Q^2}\right)+\quad\quad\quad\notag\\
+W_2(\tau)\dfrac{1}{m_N^2}\left(P^\mu-\dfrac{P.Q}{Q^2}Q^\mu\right)
\left(P^\nu-\dfrac{P.Q}{Q^2}Q^\nu\right).
\end{eqnarray}

Then the response functions in the RFG model can be written:
\begin{multline}\label{s2e8} R_{L,T}^\text{(RFG)}=\dfrac{3\mathcal{N}}{4m_N
\kappa\eta_F^3}(\varepsilon_F-\Gamma)
\Theta(\varepsilon_F-\Gamma)\times\\
\times\begin{cases} \dfrac{\kappa^2}{\tau}
\left[(1+\tau)W_{2}(\tau)-W_{1}(\tau)
+W_{2}(\tau)\Delta\right]~\text{for } L\\
\left[2W_{1}(\tau)+W_{2}(\tau)\Delta\right]~\text{for } T
\end{cases}\!\!\!\!\!\!,
\end{multline}
%where $m_N$ is the nucleon mass,
\begin{eqnarray}\label{s2e9}
W_1(\tau)=\tau
G_M^2(\tau),~W_2(\tau)=\dfrac{[G_E^2(\tau)+\tau
G_M^2(\tau)]}{1+\tau},
\end{eqnarray}
where $G_E$ and $G_M$ are the electric and magnetic Sachs form factors, the standard dimensionless variables are defined by
\begin{eqnarray}
&\kappa\equiv q/2m_N,~\lambda\equiv\omega/2m_N,~\tau=\kappa^2-\lambda^2,&\notag\\
&\eta\equiv|\mathbf{p}|/m_N,~\varepsilon\equiv E(\mathbf{p})/m_N=\sqrt{1+\eta^2},&\notag\\
&\eta_F\equiv  p_F/m_N,~\varepsilon_F=\sqrt{1+\eta_F^2},&\label{s2e10}
\end{eqnarray}
and
\begin{eqnarray}
\Delta\!&=&\!\dfrac{\tau}{\kappa^2}\left[\dfrac{1}{3}(\varepsilon_F^2
+\varepsilon_F\Gamma+\Gamma^2)+\lambda(\varepsilon_F+\Gamma)
+\lambda^2\right]\!-\!(1+\tau),\notag\\
\Gamma&\equiv&\max\left[(\varepsilon_F-2\lambda),\gamma_{\text{--}}\equiv
\kappa\sqrt{1+\dfrac{1}{\tau}}-\lambda\right].\label{s2e11}
\end{eqnarray}

%The scaling variable $\psi$ is defined by~\cite{ant01,ant02}
%\begin{eqnarray}\label{s2e12}
%\psi\equiv \frac{1}{\sqrt{\xi_F}}
%\frac{\lambda-\tau}{\sqrt{(1+\lambda)\tau+
%\kappa\sqrt{\tau(1+\tau)}}},
%\end{eqnarray}
%where $\xi_F = \sqrt{(1+ \eta_F^2)}-1$.
%%%%%%%%%%%%%%%%%%%%%%%%

In Refs.~\cite{ant18,ant19,ant20,ant21,ant22} we defined~\cite{ant18,ant19} and applied the scaling function within the CDFM using the basis of the RFG scaling function. In the model~\cite{ant15,ant16} the one-body density matrix $\rho(\mathbf{r},\mathbf{r}')$ is an infinite superposition of one-body density matrices $\rho_x(\mathbf{r},\mathbf{r}')$ corresponding to single Slater determinant wave functions of systems of free $A$ nucleons homogeneously distributed in a sphere with radius $x$, density $\rho_0(x)=\dfrac{3A}{4\pi x^3}$ and Fermi momentum $p_F(x) = \left[\dfrac{3\pi^2}{2}\rho_0(x)\right]^{1/3} = \dfrac{\alpha}{x}$ (with $\alpha \approx 1.52 A^{1/3}$):
\begin{equation}\label{s2e13}
\rho({\mathbf{r}},{\mathbf{r'}})=\int\limits_0^\infty dx
|F(x)|^2\rho_x({\mathbf{r}},{\mathbf{r'}}).
\end{equation}

The weight function $|F(x)|^2$ can be expressed in an equivalent way either by means of the density distribution~\cite{ant15,ant16,ant19}:
\begin{equation}\label{s2e14}
|{F}(x)|^{2}=-\frac{1}{\rho_0(x)} \left. \frac{d\rho(r)}{dr}\right |_{r=x}\mbox{ at }\frac{d\rho(r)}{dr}\leq0
\end{equation}
or by the nucleon momentum distribution~\cite{ant19}:
\begin{equation}\label{s2e15}
|{F}(x)|^{2}=-\frac{3\pi^{2}}{2}\frac{\alpha}{x^{5}} \left.
\frac{dn(k)}{dk}\right |_{k={\alpha}/{x}}\mbox{ at
}\frac{dn(k)}{dk}\leq0
\end{equation}

In Eqs.~(\ref{s2e14}) and (\ref{s2e15})
\begin{gather}
\int\rho(\mathbf{r})d\mathbf{r}=A,\quad\int
n(\mathbf{k})d\mathbf{k}=A,~\text{and}\notag\\
\int\limits_{0}^{\infty}|F(x)|^{2}dx=1. \label{s2e16}
\end{gather}

%In the first version of the CDFM applied to the scaling phenomenon~\cite{ant18,ant19,ant20,ant21,ant22} the RFG scaling function $f_\text{RFG}(\psi^\prime,x)$ is directly weighted by the function $|F(x)|^2$ and integration over $x$ is performed from zero to infinity.

So, in the first version of the CDFM approach the constructed (CDFM$_\text{I}$) scaling function has the form~\cite{ant18,ant19}:
\begin{equation}\label{s2e16dop1}
f(\psi')= \int\limits_{0}^{\alpha/(k_{F}|\psi'|)}dx |F(x)|^{2}
f_{RFG}(x,\psi'),
\end{equation}
where the RFG scaling function is
\begin{align}
f_{RFG}(x,\psi') =& \displaystyle \frac{3}{4} \left[\! 1\!-\!\left(\!
\frac{k_Fx|\psi'|}{\alpha}\! \right)^{2}\!\right]\! \left\{\! 1\!+\! \left(
\!\frac{xm_N}{\alpha}\!\right)^2 \!\left(\! \frac{k_Fx|\psi'|}{\alpha}
\!\right)^2 \right. \nonumber\\
& \times \displaystyle \left. \left[2+ \left( \frac{\alpha}{xm_N}
\right)^2- 2\sqrt{1+ \left( \frac{\alpha}{xm_N} \right)^2}\right]
\right\} \label{s2e16dop2}
\end{align}
and the momentum $k_F$ is calculated consistently in the CDFM for each nucleus
from the expression:
\begin{equation} \label{s2e16dop3}
k_F= \int_{0}^{\infty} dx k_{F}(x)|F(x)|^2=
\int_{0}^{\infty} dx \frac{\alpha}{x}|F(x)|^{2}.
\end{equation}
Thus, $k_F$ in CDFM is not a fitting parameter as it is in the RFG model.

Using Eqs.~(\ref{s2e14}) and~(\ref{s2e15}) in Eqs.~(\ref{s2e16dop1}) and~(\ref{s2e16dop3}) the CDFM$_\text{I}$ scaling function $f(\psi')$ and $k_F$ can be expressed explicitly by the density and momentum distributions~\cite{ant19}.

On the contrary to the CDFM$_\text{I}$, in this work we construct a more general CDFM approach (CDFM$_\text{II}$) starting not from the scaling function, but from the hadronic tensor, the response functions and related quantities in the model of the Relativistic Fermi Gas with a density $\rho_0(x)$ and a Fermi momentum $p_F(x)$. Thus, now we replace the quantities $p_F$, $\eta_F$ and $\varepsilon_F$ in Eqs.~(\ref{s2e6}), (\ref{s2e8}), and (\ref{s2e11}) by
\begin{eqnarray}
\eta_F(x)&=&\dfrac{p_F(x)}{m_N}=\dfrac{\alpha}{xm_N},\notag\\
\varepsilon_F(x)&=&\sqrt{1+\eta^2_F(x)}=\sqrt{1+\left(\dfrac{\alpha}{xm_N}\right)^2}\label{s2e17}
\end{eqnarray}
and, following the CDFM methods in~\cite{ant15,ant16}, the hadronic tensor and the response functions in the CDFM$_\text{II}$ approach are obtained by weighting the RFG model ones by the function $|F(x)|^2$  [Eqs.~(\ref{s2e14}) and (\ref{s2e15})]:
\begin{eqnarray}
W^{\mu\nu}_\text{CDFM}&=&\int\limits_{0}^{\infty}|F(x)|^2W^{\mu\nu}_\text{(RFG)}(x) dx,\label{s2e18dop}\\
R_L(\psi)&=&\int\limits_{0}^{\infty}|F(x)|^2R_{L}^\text{(RFG)}(x,\psi)dx,\label{s2e18}\\
R_T(\psi)&=&\int\limits_{0}^{\infty}|F(x)|^2R_{T}^\text{(RFG)}(x,\psi)dx,\label{s2e19}
\end{eqnarray}
where $W^{\mu\nu}_\text{(RFG)}(x)$ and $R_{L,T}^\text{(RFG)}(x,\psi)$ are given by Eq.~(\ref{s2e6}) and Eq.~(\ref{s2e8}), respectively, but now the Fermi momentum depending on $x$  according to Eq.~(\ref{s2e17}) and the scaling variable $\psi $ is defined by~\cite{ant01,ant02}
\begin{eqnarray}\label{s2e12}
\psi\equiv \frac{1}{\sqrt{\xi_F}}
\frac{\lambda-\tau}{\sqrt{(1+\lambda)\tau+
\kappa\sqrt{\tau(1+\tau)}}},
\end{eqnarray}
where $\xi_F = \sqrt{(1+ \eta_F^2)}-1$. Note that Eq.~(\ref{s2e12}) is meant to be used only in the Pauli unblocked region $q>2k_F$.

We label $\dfrac{d^2\sigma}{d\Omega d\varepsilon'}$ by
$C^\text{CDFM}(\psi)$:
\begin{eqnarray}
&C^\text{CDFM}(\psi)\equiv\dfrac{d^2\sigma}{d\Omega
d\varepsilon'}=\qquad\qquad\qquad\qquad\qquad\qquad\qquad&\notag\\
&=\sigma_M\bigg\{\!\!\left(\dfrac{Q^2}{q^2}\right)^2\!\!R_L(\psi)\!+\!\left[
\dfrac{1}{2}\left|\dfrac{Q^2}{q^2}\right|+\tan^2\dfrac{\theta}{2}\right]\!\!
R_T(\psi)\!\!\bigg\}.&\label{s2e27}
\end{eqnarray}
The single-nucleon $eN$ elastic cross section has the form~\cite{ant24}:
\begin{eqnarray}\label{s2e28}
S\!=\!\sigma_M\bigg\{\!\!\left(\dfrac{Q^2}{q^2}\right)^2\!G_L(\tau)\!+\!\left[
\dfrac{1}{2}\left|\dfrac{Q^2}{q^2}\right|\!+\!\tan^2\dfrac{\theta}{2}\right]\!
G_T(\tau)\!\!\bigg\}\!,
\end{eqnarray}
where the single-nucleon functions $G_L$ and $G_T$ are given by:
\begin{eqnarray}
&G_L(\tau)=\dfrac{\kappa}{2\tau}[ZG^2_{E,p}(\tau)+NG^2_{E,n}(\tau)]+{\cal O}({\eta}^2_F)&\label{s2e29}\\
&G_T(\tau)=\dfrac{\tau}{\kappa}[ZG^2_{M,p}(\tau)+NG^2_{M,n}(\tau)]+{\cal
O}({\eta}^2_F).&\label{s2e30}
\end{eqnarray}
Then the superscaling function can be obtained by
\begin{eqnarray}\label{s2e31}
f^\text{CDFM$_\text{II}$}(\psi)=p_F\times\dfrac{C^\text{CDFM}(\psi)}{S},
\end{eqnarray}
and, finally, following~\cite{ant04}  longitudinal $L$ and transverse $T$ scaling functions can be introduced:
\begin{eqnarray}
f_L(\psi)=p_F\times\dfrac{R_L(\psi)}{G_L},\label{s2e32}\\
f_T(\psi)=p_F\times\dfrac{R_T(\psi)}{G_T}.\label{s2e33}
\end{eqnarray}
We note that this approach differs from the first version of the CDFM applied to the scaling phenomenon~\cite{ant18,ant19,ant20,ant21,ant22} where the RFG scaling function $f_\text{RFG}(\psi^\prime,x)$ is directly weighted by the function $|F(x)|^2$ (Eqs.~(\ref{s2e16dop1}) and~(\ref{s2e16dop2})).

As mention in the Introduction, in this paper we mark the CDFM approach developed in our previous works~\cite{ant18,ant19,ant20,ant21,ant14,ant56,ant22} as CDFM$_\text{I}$ in contrast with the CDFM$_\text{II}$ one presented in this Section~\ref{sect2ant}. We would like to note that, as can be seen in Section~\ref{sect3ant}, in the CDFM$_\text{II}$: $f_L^\text{CDFM$_\text{II}$}(\psi )\neq f_T^\text{CDFM$_\text{II}$}(\psi)$ in contrast with  CDFM$_\text{I}$, where
%scaling function is defined on the basis of the scaling function of the RFG model and
$f_L^\text{CDFM$_\text{I}$}(\psi )= f_T^\text{CDFM$_\text{I}$}(\psi)$.
%Another important feature of the scaling functions ($f_L^\text{CDFM$_\text{II}$}(\psi)$, $f_T^\text{CDFM$_\text{II}$}(\psi)$, $f^\text{CDFM$_\text{II}$}(\psi)$) in the CDFM$_\text{II}$ model is that they can be calculated for fixed values of momentum transfer $q$, not only for $q>2p_F$, but also for $q<2p_F$ (where scaling does not hold in the RFG model).
The results and discussions are given in the next Section~\ref{sect3ant}.

\section[]{RESULTS\label{sect3ant}}

\begin{figure*}[t]
\centering
\includegraphics[width=172mm]{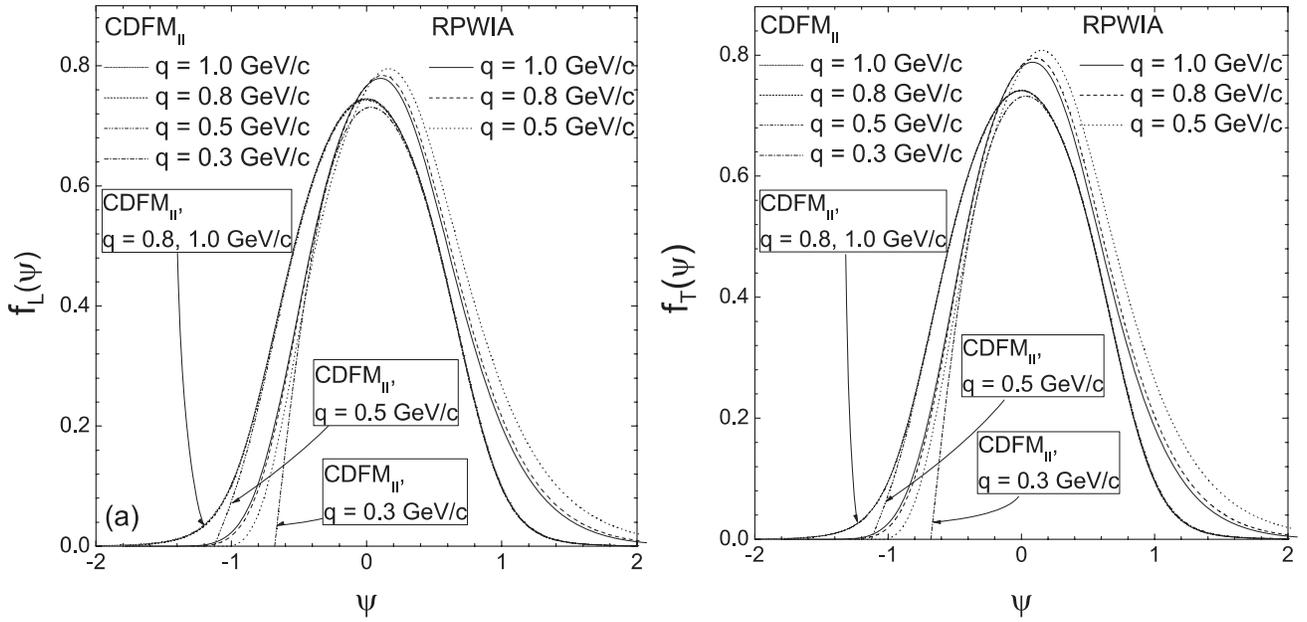}
\caption[]{The longitudinal scaling functions $f_L(\psi)$ (a) and  the transverse scaling functions $f_T(\psi)$ (b) for $^{12}$C calculated in the CDFM$_\text{II}$ for $q=0.3$, $0.5$, $0.8$, and $1.0$~GeV/c and RPWIA (Lorentz gauge) for $q=0.5$, $0.8$, and $1.0$~GeV/c.\label{fig1ant}}
\end{figure*}

\begin{figure}[t]
\centering
\includegraphics[width=86mm]{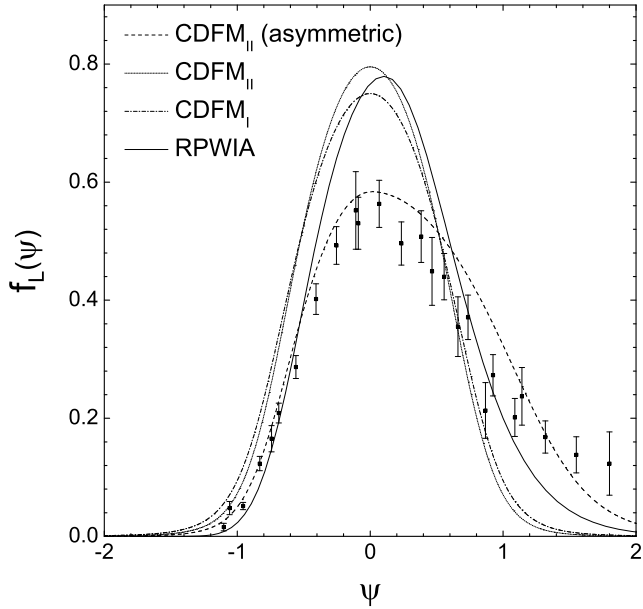}
\caption[]{The longitudinal scaling functions $f_L(\psi)$ for $^{12}$C calculated for $q=1$~GeV/c in the CDFM$_\text{I}$, CDFM$_\text{II}$, RPWIA and asymmetric CDFM$_\text{II}$. The experimental data are taken from~\cite{ant25}.\label{fig2ant}}
\end{figure}

\begin{figure}[t]
\centering
\includegraphics[width=86mm]{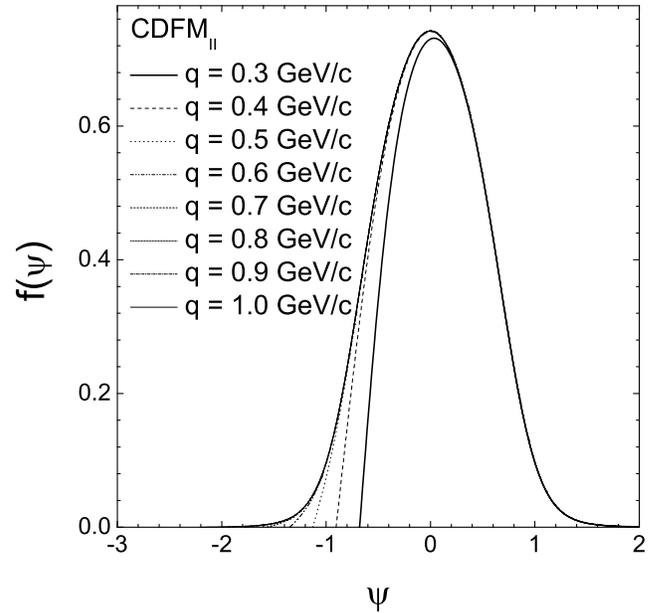}
\caption[]{The quasielastic scaling function $f^\text{QE}(\psi)$ for $^{12}$C calculated in the CDFM$_\text{II}$ for $q=0.3-1.0$~GeV/c with step $0.1$~GeV/c.\label{fig3ant}}
\end{figure}
\begin{figure}[t]
\centering
\includegraphics[width=86mm]{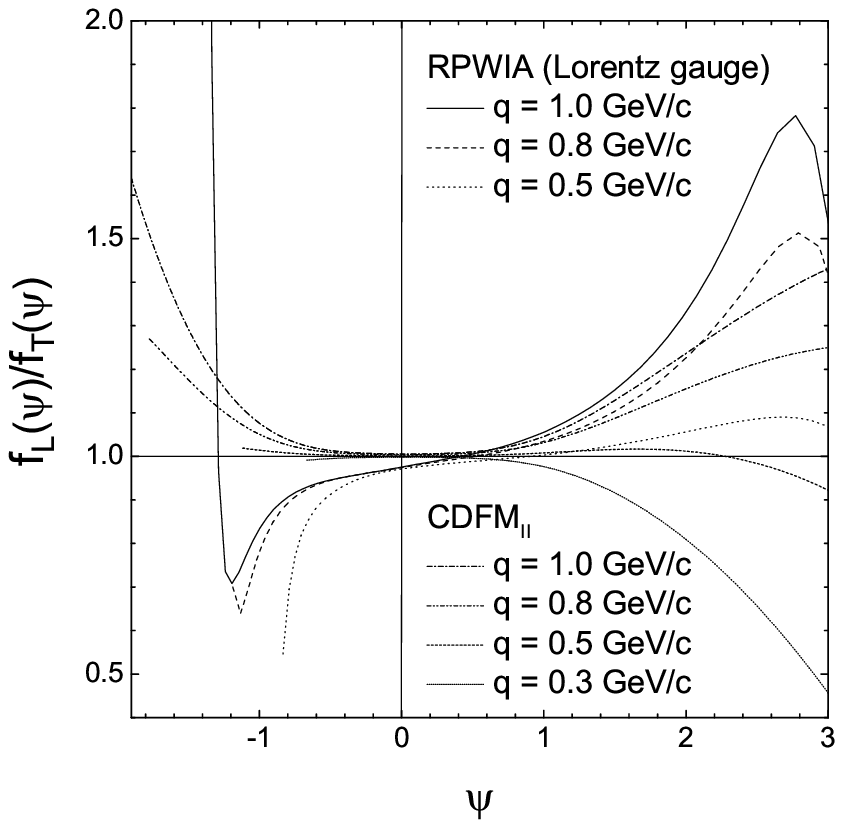}
\caption[]{The ratio $f_L(\psi)/f_T(\psi)$ for $^{12}$C calculated in
the CDFM$_\text{II}$ for $q=0.3$, $0.5$, $0.8$, and $1.0$~GeV/c and RPWIA (Lorentz gauge) for $q=0.5$, $0.8$, and $1.0$~GeV/c.\label{fig4ant}}
\end{figure}
\begin{figure}[t]
\centering
\includegraphics[width=86mm]{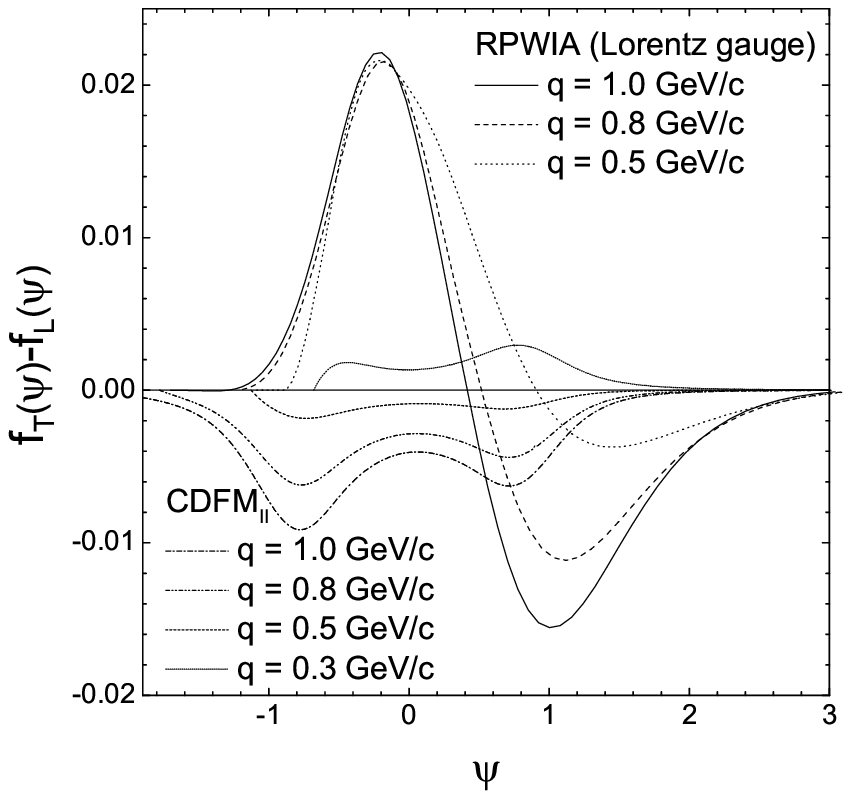}
\caption[]{The differences $f_T(\psi)-f_L(\psi)$ for $^{12}$C calculated in
the CDFM$_\text{II}$ for $q=0.3$, $0.5$, $0.8$, and $1.0$~GeV/c and RPWIA (Lorentz gauge) for $q=0.5$, $0.8$, and $1.0$~GeV/c.\label{fig5ant}}
\end{figure}

\begin{figure*}[t]
\centering
\includegraphics[width=170mm]{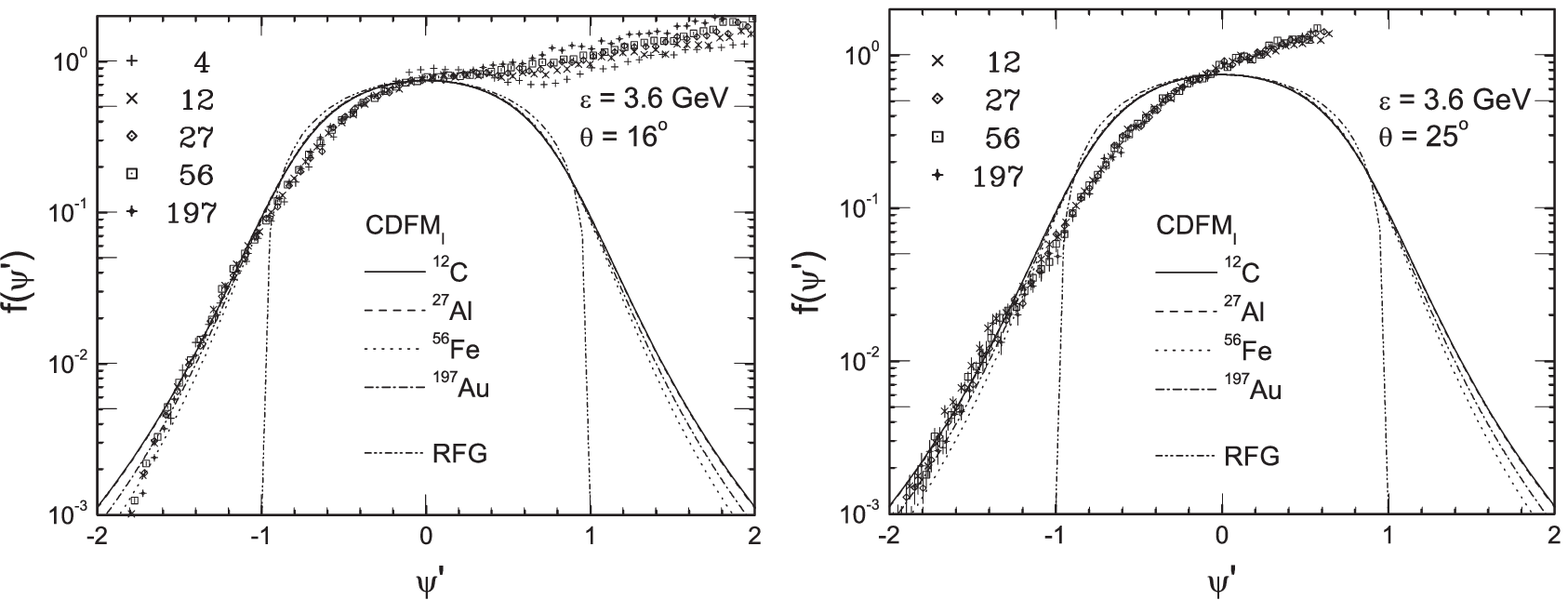}
\caption[]{The quasielastic scaling function $f^\text{QE}(\psi^\prime)$ for $^{12}$C, $^{27}$Al, $^{56}$Fe, and $^{197}$Au calculated in the CDFM$_\text{I}$ and RFG. The experimental data are taken from~\cite{ant03} and the labels indicate the mass number for each set of data.\label{fig6aant}}
\end{figure*}

\begin{figure*}[t]
\centering
\includegraphics[width=170mm]{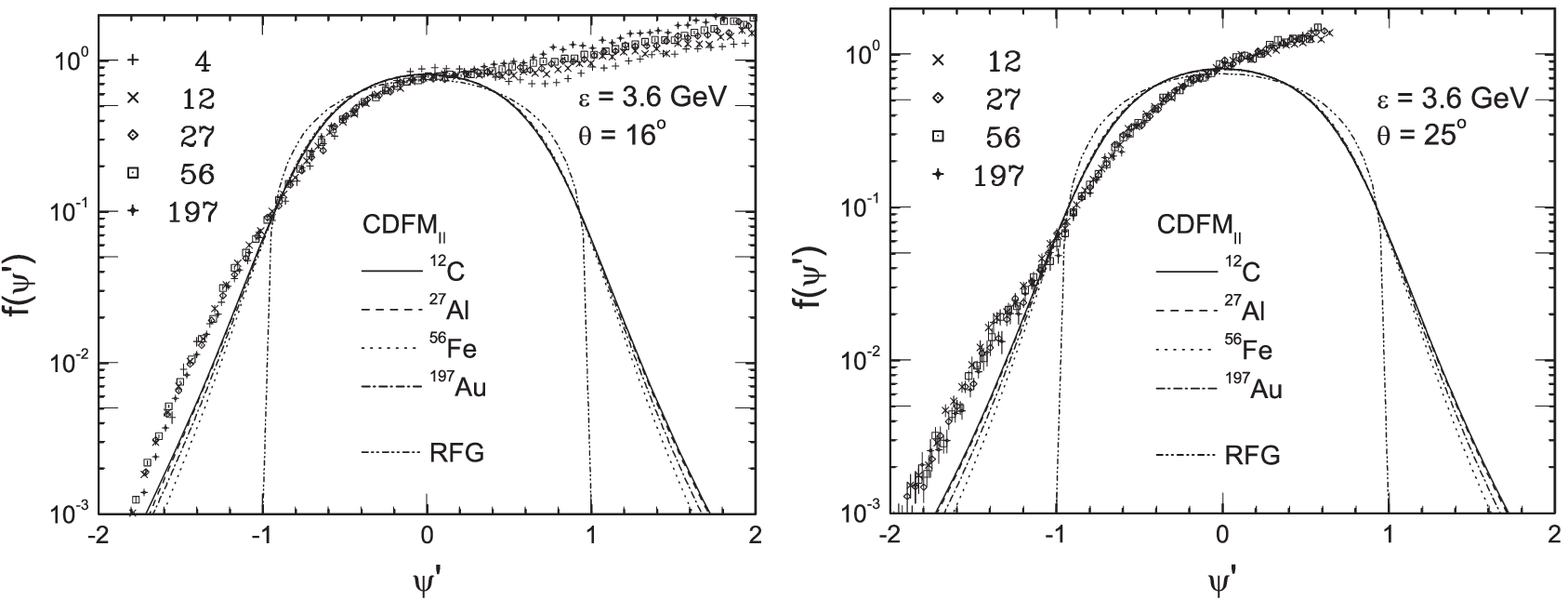}
\caption[]{The quasielastic scaling function $f^\text{QE}(\psi^\prime)$ for $^{12}$C, $^{27}$Al, $^{56}$Fe, and $^{197}$Au calculated in the CDFM$_\text{II}$ and RFG. The experimental data are taken from~\cite{ant03} and the labels indicate the mass number for each set of data.\label{fig6bant}}
\end{figure*}

\begin{figure}[t]
\centering
\includegraphics[width=86mm]{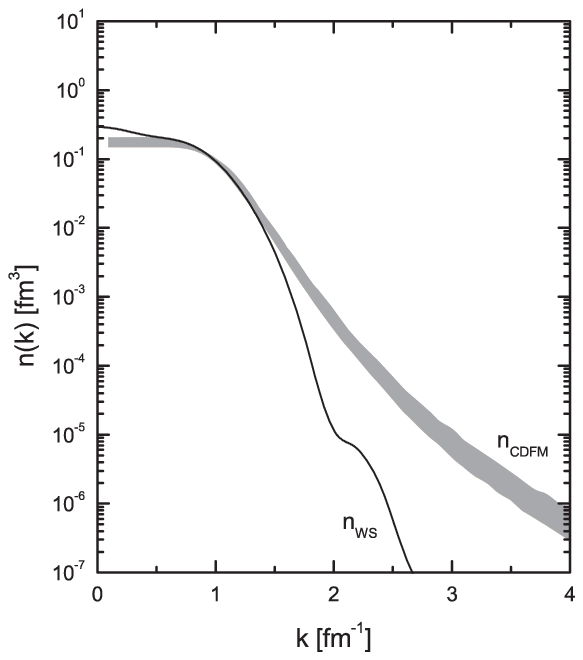}
\caption[]{The nucleon momentum distribution $n(k)$ (also see Fig.~\ref{fig2ant} of~\cite{ant19} and Fig.~\ref{fig3ant} of~\cite{ant21}). Gray area: combined results of CDFM for $^4$He, $^{12}$C, $^{27}$Al, $^{56}$Fe and $^{197}$Au. Solid line: mean-field result using Woods-Saxon single-particle wave functions (for $^{56}$Fe). The normalization is: $\int n(k)d^3k = 1$.\label{fignew}}
\end{figure}

In this section we present firstly our results of longitudinal (Fig.~\ref{fig1ant}a) and transverse (Fig.~\ref{fig1ant}b) scaling functions at fixed values of momentum transfer $q=0.3$, $0.5$, $0.8$ and $1.0$~GeV/c calculated within the CDFM$_\text{II}$ approach  compared with results of the relativistic plane wave impulse approximation (RPWIA) with Lorentz gauge~\cite{ant37}. In contrast with our previous results, where the CDFM$_\text{I}$ scaling functions are equal, $f^\text{CDFM$_\text{I}$}_L(\psi)=f^\text{CDFM$_\text{I}$}_T(\psi)=f^\text{CDFM$_\text{I}$}(\psi)$, and do not depend on the momentum transfer $q$, in the CDFM$_\text{II}$ the scaling functions, which are built from the nuclear electromagnetic response functions, depend on the momentum transfer $q$ till a sufficiently high $q$. As can be seen from Fig.~\ref{fig1ant}, scaling of first kind is clearly violated for low $q$-values ($q<0.5$~GeV/c) in the negative $\psi$-region, whereas for $q$ of the order of $0.5$~GeV/c, scaling violation slowly disappears as $q$ increases and the CDFM$_\text{II}$ and RPWIA scaling functions reach their asymptotic values. In Fig.~\ref{fig2ant} we give the comparison of the longitudinal scaling functions from CDFM$_\text{I}$, CDFM$_\text{II}$ and RPWIA with the experimental data. We note that all the three approaches overestimate the data, especially at the QE peak and in the high positive $\psi$ region. A better comparison with data can be achieved by introducing, as done in our previous work~\cite{ant21}, a phenomenological asymmetric tail for $\psi>0$ in such a way to preserve the correct normalization of the superscaling function: the corresponding result, labelled by ``CDFM$_\text{II}$ (asymmetric)'', is also shown in the figure. Similar behavior for the total quasielastic scaling function can be seen in the next Fig.~\ref{fig3ant}, where we present our results of calculations of $f^\text{QE}(\psi)$ [Eq.~(\ref{s2e31})] for $^{12}$C within the CDFM$_\text{II}$ model for $q=0.3-1.0$~GeV/c with a step of $0.1$~GeV/c. Note that the asymmetry in the scaling function, clearly observed for low-$q$ values, tends to disappear as $q$ goes up. This asymmetry in the negative $\psi$-region at low $q$ is linked to effects introduced by Pauli-blocking which destroy the scaling behavior. We note that the results of CDFM$_\text{II}$ at  $q=0.8$~GeV/c and  $q=1.0$~GeV/c are similar among themselves in both the negative and positive $\psi$-regions.

In Figs.~\ref{fig4ant} and~\ref{fig5ant} we present results for the ratio $f_L(\psi)/f_T(\psi)$ and the differences $f_T(\psi)-f_L(\psi)$ for $^{12}$C calculated in the CDFM$_\text{II}$ and RPWIA (Lorentz gauge) at fixed values of momentum transfer $q=0.3$, $0.5$, $0.8$, and $1.0$~GeV/c. In the CDFM$_\text{II}$ calculations we observe violation of the scaling of 0$^\text{th}$ kind ($f_L(\psi)\neq f_T(\psi)$), at variance with the CDFM$_\text{I}$ one. The behavior of the ratio $f_L(\psi)/f_T(\psi)$ in our model is similar to that in the RPWIA for positive $\psi$-values where the response is positive except for very low-$q$ ($q=0.3$~GeV/c). On the contrary, in the negative $\psi$-region, the ratio $f_L(\psi)/f_T(\psi)$ becomes negative for RPWIA and positive for CDFM$_\text{II}$, being the variation in the former case much larger. These results are consistent with the ones shown in Fig.~\ref{fig5ant}. Here, the difference $f_T(\psi)-f_L(\psi)$ is negative (positive) for all $q$-values ($q=0.3$~GeV/c) in the whole $\psi$-region in the case of the CDFM$_\text{II}$ model. This is in contrast with RPWIA results where $f_T(\psi)-f_L(\psi)$ starts being positive (left $\psi$-region) changing to negative for higher $\psi$. The specific value of $\psi$ where $f_T(\psi)-f_L(\psi)$ changes sign depends on the $q$-value considered being larger as $q$ increases. As a general outcome, we conclude by observing that CDFM$_\text{II}$ scaling functions are not so different from each other as they are in the RPWIA case.

%The behavior of the ratio $f_L(\psi)/f_T(\psi)$ in our model is similar with that in the RPWIA, in contrast with the differences $f_T(\psi)-f_L(\psi)$, where our results for $q=300$~MeV/c are positive, but for $q=500$, $800$, and $1000$~MeV/c are negative. We note that, generally, our CDFM$_\text{II}$ scaling functions are not so different among themselves as they are in the RPWIA case.

Next step in our studies is to examine the scaling of the second kind in the CDFM$_\text{II}$. This requires calculations of the scaling functions for different nuclei. In Figs.~\ref{fig6aant} and~\ref{fig6bant} we give the results for the quasielastic scaling functions for $^{12}$C, $^{27}$Al, $^{56}$Fe, and $^{197}$Au calculated in the CDFM$_\text{I}$ and CDFM$_\text{II}$, respectively. The result of the RFG model is also presented. One can see the essential difference between the results of the RFG model and those of the CDFM$_\text{I}$ and CDFM$_\text{II}$ in the region $\psi'<-1$. For readers who may not be familiar with the scaling variable $\psi'$ we recall that first the variable $\psi$ was introduced by W.M. Alberico \emph{et al.} \cite{ant01} as the natural scaling variable within RFG model. $\psi$ is defined (see Eq.~(\ref{s2e12})) so that it varies from $\psi\sim (-\kappa/\sqrt{\xi_F})$ to $\psi=0$ in the left hand side of the quasielastic peak, \emph{i.e.}, when the transfer energy $\omega$ varies from $0$ to $Q^2/2m_N$, while $\psi>0$ when we are in the right hand side and other production channels may start to open. The variable $\psi'$ was introduced in~\cite{ant11} and \cite{ant03,ant04} as the corresponding phenomenological variable to analyse data and to show scaling of second kind. It involves a redefinition of $\lambda$ that corrects for the displacement in $\omega$ of the quasielastic peak position, which depends on the nuclear target. It can be seen also from our results that the scaling of the second kind is good in both CDFM approaches, however, the CDFM$_\text{I}$ scaling functions are in better agreement with the experimental data. This is due to the fact that the maximum of $f^\text{CDFM$_\text{II}$}(\psi)$ is $0.80$ (coming from the expressions for the RFG hadronic tensor) but not $0.75$ as it is in the RFG and, correspondingly, in the CDFM$_\text{I}$ . In this case:
\[
f^\text{CDFM$_\text{II}$}_{\max}(\psi)\approx0.8>f^\text{CDFM$_\text{I}$}_{\max}(\psi)=0.75
\]
and the normalization
\[
\int\limits_{-\infty}^\infty f^\text{CDFM$_\text{II}$}(\psi)d\psi=\int\limits_{-\infty}^\infty f^\text{CDFM$_\text{I}$}(\psi)d\psi=1
\]
leads to narrower behavior of $f(\psi)$ in the CDFM$_\text{II}$.

The behavior of the CDFM$_\text{I}$ and CDFM$_\text{II}$ scaling functions can be explained by the long-range collective correlations included in the CDFM which is based on the Generator Coordinate Method~\cite{ant17} applied to consider the monopole breathing motions~\cite{ant15,ant16}. These correlations are important and they are reflected in the tail of the CDFM scaling functions at negative $\psi'$. On the contrary, the results of mean-field approaches (relativistic or not) are generally closer to those of the RFG model. Although the differences of the results of CDFM$_\text{I}$ and CDFM$_\text{II}$
are not large, they reflect the different stage at which the RFG approach is replaced by the CDFM (using the weight function): in the CDFM$_\text{II}$ long-range correlations are included at the level of the hadron tensor, whereas in the CDFM$_\text{I}$ this is done directly in the scaling function after having factorized and divided by the single-nucleon factors.

%The behavior of the CDFM$_\text{I}$ and CDFM$_\text{II}$ scaling functions at $\psi'<-1$ can be explained by the correlations included in the CDFM. The latter are important in this $\psi'$-region, while the results of mean-field approaches (relativistic or not) are generally closer to that of the RFG model than the CDFM's are.

In order to illustrate the effects of the NN correlations included in the CDFM on the tail of the CDFM scaling function we remind here the relationship (mentioned in the Introduction) between the scaling function $f(\psi')$ and the nucleon momentum distribution $n(k)$. It was found within the CDFM~\cite{ant14,ant18,ant19,ant21} that the slope of the QE scaling function  $f(\psi')$ at negative $\psi'$ crucially depends on the high-momentum tail of $n(k)$ at larger values of $k$ ($k> 1.5$~[fm$^{-1}$]). It can be seen in Fig.~\ref{fignew} the difference between the CDFM combined results ($n_\text{CDFM}$) for $^4$He, $^{12}$C, $^{27}$Al, $^{56}$Fe and $^{197}$Au (gray area) and the mean-field result ($n_\text{WS}$) obtained by means of Woods-Saxon single-particle wave functions (for $^{56}$Fe). It was shown in~\cite{ant18,ant19} that \emph{when the scaling function is calculated using realistic high-momentum components of $n(k)$ at} $k> 1.5$~[fm$^{-1}$] (\emph{i.e.} obtained in a nuclear model accounting for NN correlations beyond the mean-field approximation), \emph{a reasonable explanation of the superscaling behavior of the scaling function for $\psi'< -1$ is achieved}. We note that the difference between the CDFM scaling function and that from the RFG model for $|\psi'| > 1$  which can be seen in Figs.~\ref{fig6aant} and~\ref{fig6bant} is due to the large difference between $n(k)$ in CDFM and that in the RFG model, where the (dimensionless) momentum distribution is a step function. The study performed in~\cite{ant19} of the sensitivity of the CDFM scaling function to the particular behavior of $n(k)$ in different regions of $k$ showed that the available empirical data on $f(\psi')$ are informative for $n(k)$ for momentum $k \leq 2.0$--$2.5$~[fm$^{-1}$].

\begin{figure*}[p]
\centering
\includegraphics[width=155mm]{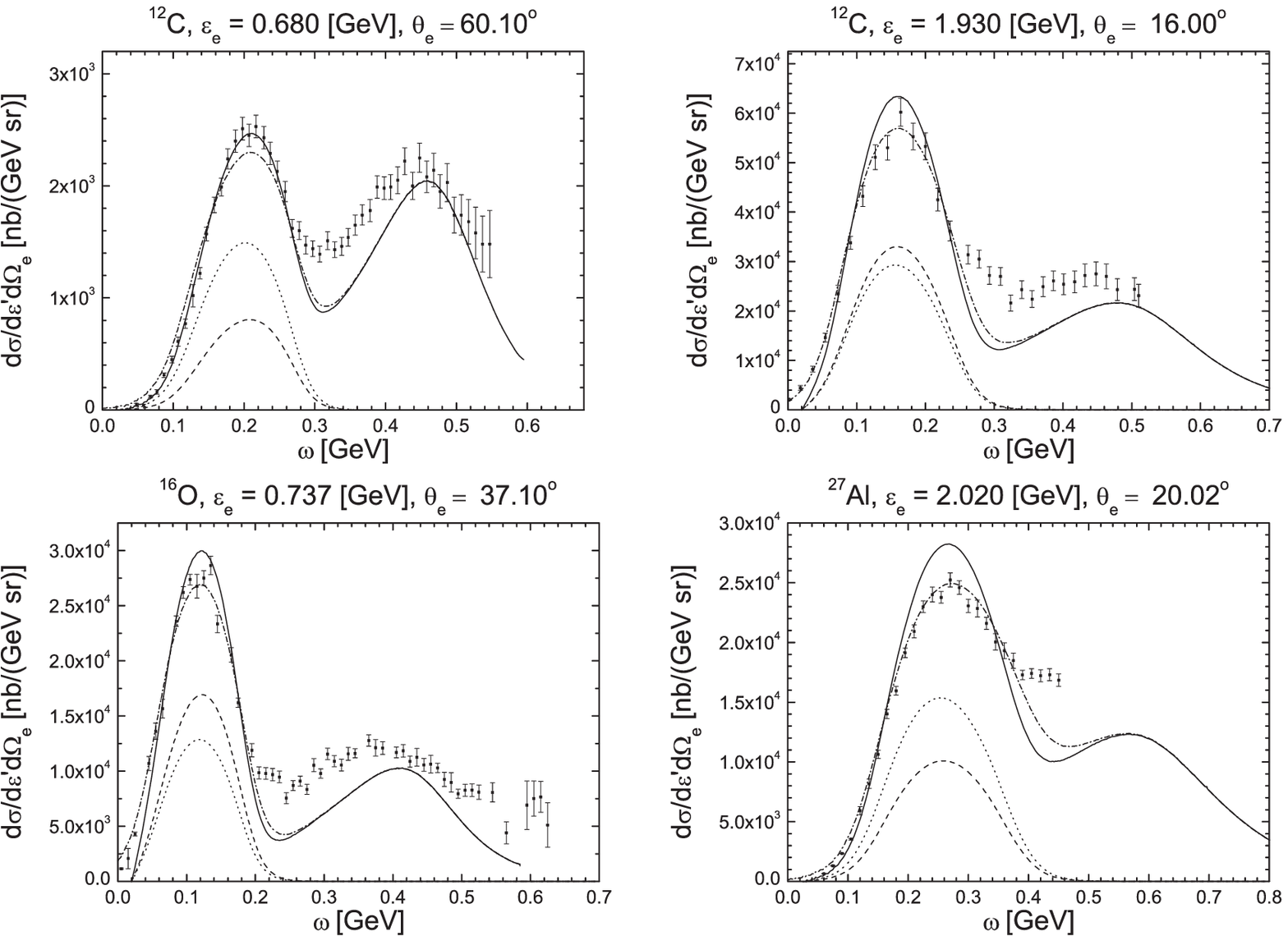}
%\end{figure*}
%
%\begin{figure*}[t]
%\centering
\includegraphics[width=155mm]{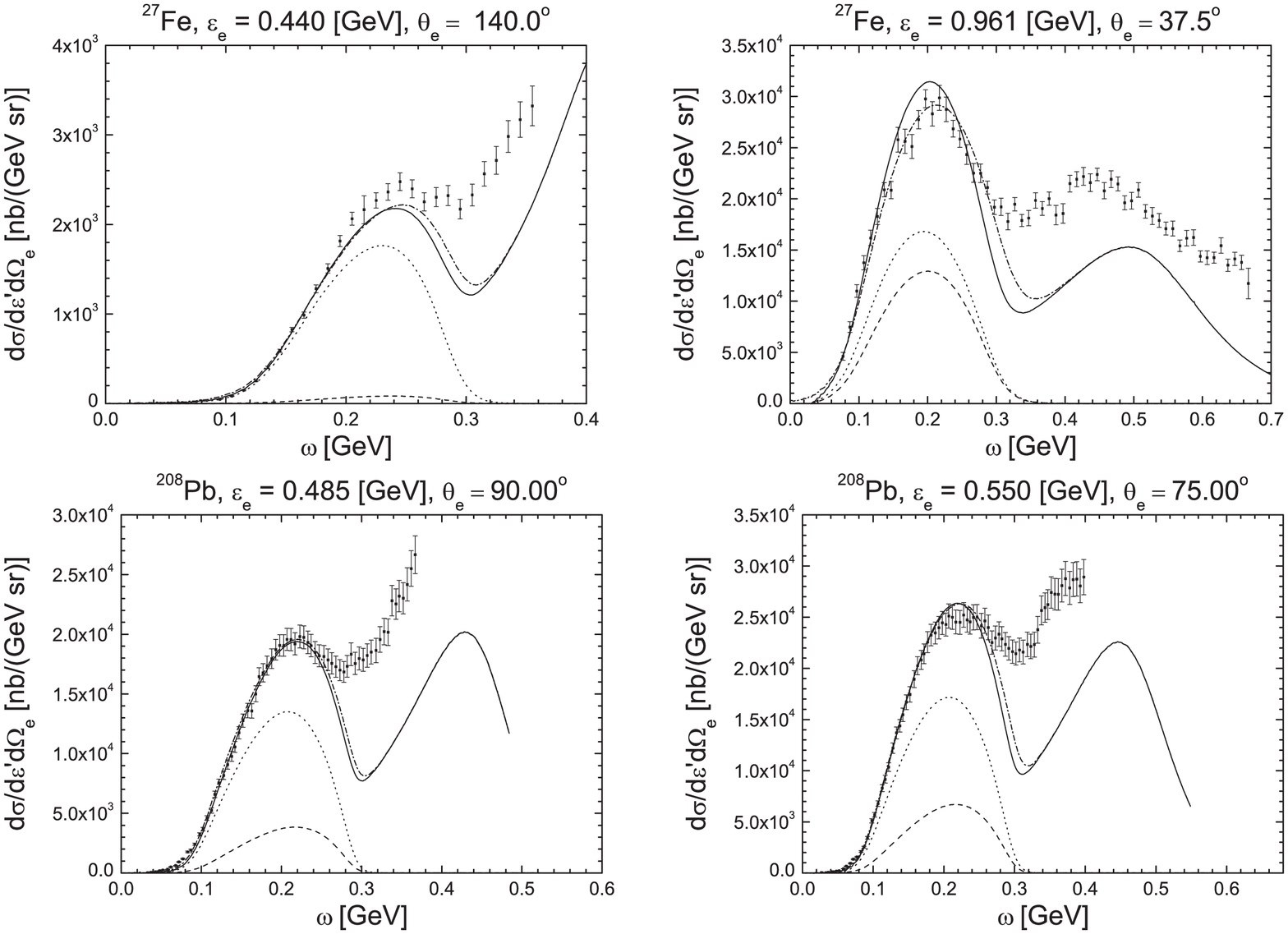}
\caption[]{Inclusive electron cross sections as function of energy loss. The results are given: the CDFM$_\text{I}$ dash-dotted line, the CDFM$_\text{II}$ solid line, the L-contribution in CDFM$_\text{II}$ dashed line, the T-contribution in CDFM$_\text{II}$ dotted line. The experimental data are taken from~\cite{ant57}.\label{fig7ant}}
\end{figure*}

A test of the CDFM superscaling functions is performed (Fig.~\ref{fig7ant}) by calculations of the cross sections of electron scattering in quasielastic and $\Delta$ region for nuclei with $12\leq A\leq208$ at different energies and angles using the CDFM$_\text{I}$ and CDFM$_\text{II}$ scaling functions. For the scaling function in the $\Delta $ region we use the results of Ref.~\cite{ant19}. The results are compared with available experimental data.

As can be seen from Fig.~\ref{fig7ant} the results calculated with both CDFM$_\text{I}$ and CDFM$_\text{II}$ scaling functions do not differ too much, agreeing well with experimental data in the QE region. In some particular cases, CDFM$_\text{II}$ overestimates data whereas CDFM$_\text{I}$ agrees better, being the reverse in other situations. Finally, some kinematical regimes lead to very similar results for both models, being in excellent accord with data. Away from the QE and $\Delta$ peaks the behavior of the cross sections is due to higher resonances and, as can be expected, in some cases our results are not in good agreement with the experimental data. We also display the separate longitudinal and transverse contributions to the QE peak.

\begin{figure*}[t]
\centering
\includegraphics[width=160mm]{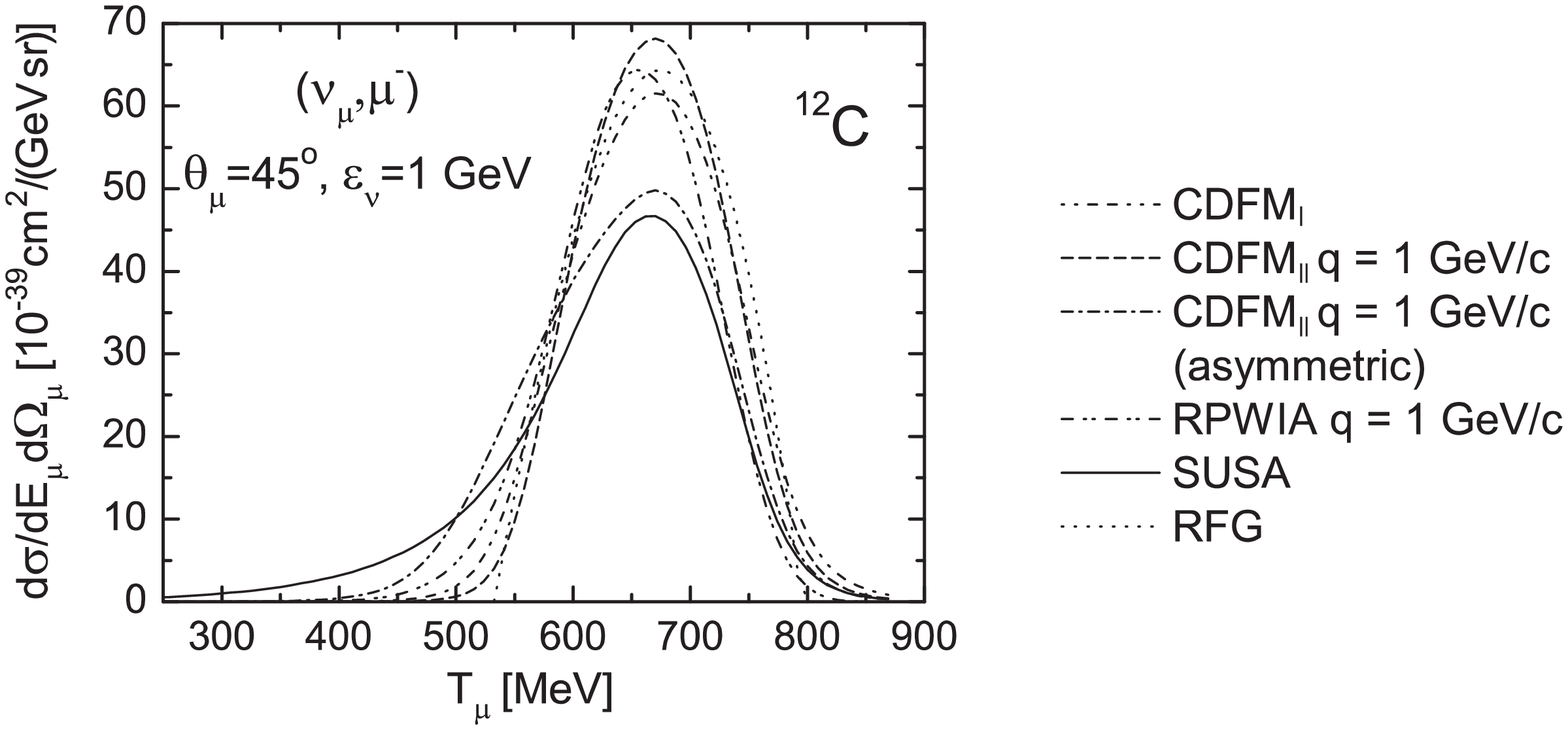}
\caption[]{The cross section of charge-changing neutrino
($\nu_\mu$,$\mu^{-}$) reaction on $^{12}$C at
$\theta_\mu=45^\circ$ and $\varepsilon_{{\nu}}=1$~GeV.\label{fig8ant}}
\end{figure*}

Finally, in Fig.~\ref{fig8ant} we present the CDFM results for the cross section of the charge-changing neutrino ($\nu_\mu$,$\mu^{-}$) reaction on $^{12}$C at $\theta_\mu=45^\circ$ and $\varepsilon_{{\nu}}=1$~GeV. The calculations are performed following the formalism from~\cite{ant21,ant24} using not only the CDFM$_\text{I}$  but also the CDFM$_\text{II}$ quasielastic scaling function. Our results are compared with those from RFG, SuSA and RPWIA.

We note that the CDFM results are qualitatively similar to that of Fig.~\ref{fig2ant}, namely, the result for the CDFM$_\text{II}$ with asymmetry is closer to that calculated using the phenomenological (SuSA) scaling function that is extracted from the experimental data on inclusive electron scattering. On the other hand, CDFM$_\text{I}$ and CDFM$_\text{II}$ models lead to very close results being the maximum of the scaling function slightly higher in the latter. The scaling functions for both approaches follow closely the behavior exhibited by the RPWIA one.

\section[]{CONCLUSIONS\label{sect4ant}}

The results of the present work can be summarized as follows:
\begin{enumerate}

\item A new, more general, approach within the Coherent Density Fluctuation Model is proposed (CDFM$_\text{II}$). We apply it to calculate the total $f(\psi^\prime)$, the longitudinal $f_L(\psi^\prime)$ and the transverse $f_T(\psi^\prime)$ scaling functions taking as starting point the hadronic tensor and the longitudinal and transverse response functions in the RFG model.

\item The approach leads to a slight violation of the zero-kind scaling ($f_L(\psi^\prime)\neq f_T(\psi^\prime)$) in contrast with the situation in the RFG and CDFM$_\text{I}$ models.

\item It is found that the ratio $f_L(\psi^\prime) / f_T(\psi^\prime)$ in the CDFM$_\text{II}$ has similarities with that from the RPWIA approach (with Lorentz gauge) for $\psi'$-positive.

\item It is shown that the CDFM$_\text{II}$ scaling functions calculated for different values of the transferred momentum $q$ show a saturation of its asymptotic behavior. The scaling of first kind appears at $q$ larger than $\approx 0.5$~GeV/c.

\item The CDFM scaling functions are applied to calculate cross sections of inclusive electron scattering (and their longitudinal and transverse components) in the quasielastic and $\Delta $-region for nuclei with $12\leq A\leq208$ at different energies and angles. The results are in good agreement with available experimental data, especially in the QE region.

\item The CDFM$_\text{II}$ approach is applied to calculate charge-changing neutrino (antineutrino) scattering on $^{12}$C at 1~GeV incident energy. The results are compared with those from the RFG model, as well as from the SuSA and RPWIA approaches.
\end{enumerate}

\begin{acknowledgments}
This work was partly supported by the Bulgarian National Science Fund under contracts nos~DO~02--285 and $\Phi$--1501 and by Ministerio de Educaci\'on y Ciencia (Spain) under contract nos.~FPA2006-13807-C02-01, FIS2005-01105, FIS2005-00640, FIS2008-04189, and PCI2006-A7-0548, and the Spanish Consolider-Ingenio 2010 programme CPAN (CSD2007-00042). This work is also partially supported by the EU program ILIAS N6 ENTApP WP1. M.V.I. acknowledges support from the European Operational programm HRD through contract BGO051PO001/07/3.3-02/53 with the Bulgarian Ministry of Education. M.B.B. and J.A.C. acknowledge support from the INFN-MEC agreement, project ``Study of relativistic dynamics in neutrino and electron scattering''.

\end{acknowledgments}

\end{document}